\documentclass[twocolumn]{aastex631}

\usepackage{natbib}
\usepackage{amsmath}
\usepackage{braket}
\usepackage{ulem}
\usepackage{soul}
\usepackage{color}

\newcommand{\HI}[2]{\textcolor{black}{#2}} 

\shortauthors{Kuniyoshi, Shoda, Iijima and Yokoyama}

\graphicspath{{./}{figures/}}

\begin{document}

\title{
Magnetic Tornado Properties: A Substantial Contribution to the Solar Coronal Heating via Efficient Energy Transfer
}

\author[0000-0003-1134-2770]{Hidetaka Kuniyoshi}
\affiliation{
Department of Earth and Planetary Science, The University of Tokyo, 7-3-1 Hongo, Bunkyo-ku, Tokyo 113-0033, Japan
}

\author[0000-0002-7136-8190]{Munehito Shoda}
\affiliation{
Department of Earth and Planetary Science, The University of Tokyo, 7-3-1 Hongo, Bunkyo-ku, Tokyo 113-0033, Japan
}

\author[0000-0002-1007-181X]{Haruhisa Iijima}
\affiliation{
Division for Integrated Studies, Institute for Space-Earth Environmental Research, Nagoya University, Furocho, Chikusa-ku, Nagoya, Aichi 464-8601
}
\affiliation{
Institute for Advanced Research, Institute for Space-Earth Environmental Research, Nagoya University, Furocho, Chikusa-ku, Nagoya, Aichi 464-8601
}

\author[0000-0001-5457-4999]{Takaaki Yokoyama}
\affiliation{
Astronomical Observatory, Kyoto University, Sakyo-ku, Kyoto, 606-8502, Japan
}

\begin{abstract}
In solving the solar coronal heating problem, it is crucial to comprehend the mechanisms by which energy is conveyed from the photosphere to the corona.
Recently, magnetic tornadoes, characterized as coherent, rotating magnetic field structures extending from the photosphere to the corona, have drawn growing interest as a possible means of efficient energy transfer.
Despite its acknowledged importance, the underlying physics of magnetic tornadoes remains still elusive.
In this study, we conduct a three-dimensional radiative magnetohydrodynamic simulation that encompasses the upper convective layer and extends into the corona, with a view to investigating how magnetic tornadoes are generated and efficiently transfer energy into the corona.
We find that a single event of magnetic flux concentration merger on the photosphere gives rise to the formation of a single magnetic tornado. \deleted{, with its rotational direction undergoing a single change in orientation to become opposite.}
The Poynting flux transferred into the corona is found to be four times greater in the presence of the magnetic tornado, as compared to its absence. \replaced{, as compared to its absence, due to a reduction in energy loss in the chromosphere.}{This increase is attributed to a reduction in energy loss in the chromosphere, resulting from the weakened magnetic energy cascade.}
Based on an evaluation of the fraction of the merging events, our results suggest that magnetic tornadoes contribute approximately $50 \%$ of \replaced{the coronal energy in the quiet Sun, and potentially even more in more magnetically active stars, such as M dwarfs and young solar analogs.}{the Poynting flux into the corona in regions where the coronal magnetic field strength is 10 G. Potentially, the contribution could be even greater in areas with a stronger coronal magnetic field.
} 
\end{abstract}


\section{Introduction} 
\label{sec:intro}
Magnetic heating of plasmas plays a pivotal role in a multitude of astronomical phenomena, ranging from the X-ray radiation emitted by accretion disks around black holes \citep{Galeev_1979_ApJ,Beloborodov_2017_ApJ} to the cosmic reionization process \citep{Washinoue_2019_ApJ,Washinoue_2021_MNRAS}. The heating of the solar corona \citep{Grotrian_1939,Edlen_1943_ZAP} is an example the astrophysical phenomena caused by magnetic field, which has been studied as one of the most important problems in astronomy \citep[e.g.,][]{Klimchuk_2006_SoPh,Cranmer_2019_ARAA}. The existence of the corona is not unique to the Sun; million-kelvin atmosphere is found on general low-mass main-sequence stars \citep{Pallavicini_1981_ApJ,Gudel_1997_ApJ,Ribas_2005_ApJ,Johnstone_2015_AA}. 
Due to its high temperature, the stellar corona is the dominant source of the high-energy photons that significantly affect the evolution of the planetary atmosphere \citep{Sanz-Forcada_2011_AA,Cuntz_2016_ApJ,Airapetian_2021_ApJ}. Solving the coronal heating problem thus has vital implications in the context of planetary science as well as stellar physics.

In solving the coronal heating problem, the following three issues need to be addressed: 1. energy transfer to the corona, 2. (magnetic) energy dissipation, and 3. coronal thermal response to heating \citep{Klimchuk_2006_SoPh}.  In terms of energy transfer, the passive advection of the photospheric magnetic field is the promising source of the upward Poynting flux \citep{Spruit_1981_AA,Steiner_1998_ApJ,Fujimura_2009_ApJ}, a fraction of which transmits into the corona.
Indeed, the magnetic flux concentrations \citep[MCs,][]{Muller_1983_SoPh,Solanki_1993_SSRv,Berger_1995_ApJ,Berger_2001_ApJ} on the photosphere are found to generate a sufficient amount of energy flux to heat at least quiet Sun corona \citep{Choudhuri_1993_SoPh,Choudhuri_1993_ApJ}. 
We note that flux emergence is an alternative candidate of coronal heat source \citep{Schrijver_1998_Natur,Wang_2020_ApJ,Wang_2022_SolPhys}.
For energy dissipation, several mechanisms are proposed, including magnetic-field braiding \citep{Parker_1972_ApJ,Parker_1983_ApJ,Parker_1988_ApJ,Sturrock_1981_ApJ,Berger_1991_AA}, resonant absorption \citep{Erdelyi_1995_AA,Terradas_2010_AA,Goossens_2011_SSRv}, phase mixing \citep{Heyvaerts_1983_AA,DeMoortel_2000_AA,Goossens_2012_ApJ}, shock formation \citep{Moriyasu_2004_ApJ,Antolin_2008_ApJ,Schiff_2016_ApJ} and turbulence \citep{vanBallegooijen_2011_ApJ,vanBallegooijen_2014_ApJ,Matsumoto_2018_MNRAS,Shoda_2021_AA}. From the observational point of view, coronal heating is often discussed in terms of thermal response, in particular the differential emission measure \citep[e.g.,][]{Aschwanden_2002_ApJ,Warren_2012_ApJ,Shoda_2021_AA}.
Recently, large-scale three-dimensional simulations spanning from the convection zone to the corona have been conducted \citep[e.g.,][]{Amari_2015_Nature,Hansteen_2015_ApJ,Rempel_2017_ApJ,Chen_2022_ApJ,Finley_2022_AA,Robinson_2022_arXiv}, which are often used to synthesize observational signatures \citep{Peter_2015_RSPTA,Chen_2021_AA,Breu_2022_AA,Malanushenko_2022_ApJ}. 
Whereas, in exchange for reality, these simulations are often too complicated to reveal the underlying physics. 
It is therefore still meaningful to conduct a simplified simulation that focuses on each building block. 

This work aims to improve our understanding of the energy transfer to the corona. Classically, energy is thought to be transferred by magnetohydrodynamic (MHD) waves generated on the photosphere \citep{Alfven_1947_MNRAS,Osterbrock_1961_ApJ,Taroyan_2009_SSRv}.
In particular, transverse (Alfv\'en and kink) wave is likely to experience less diffusion and refraction \citep{Stein_1972_ApJ,Narain_1996_SSRv,Cranmer_2007_ApJS,Priest_2014,Matsumoto_2014_MNRAS}, and thus, is one of the promising mechanisms of the energy transfer.
Indeed, the transverse waves are found to exhibit a significant amount of energy flux in the solar chromosphere \citep{DePontieu_2007_Science,Srivastava_2017_SciRep} and corona \citep{Hassler_1990_ApJ,Banerjee_1998_AA,Banerjee_2009_AA,McIntosh_2011_Nature,Hahn_2013_ApJ}, although whether it is sufficient for the coronal heating is controversial \citep{Tomczyk_2007_Science,Okamoto_2011_ApJ,Thurgood_2014_ApJ}.
In considering the wave propagation into the corona, we often assume that the horizontal random buffeting motion of photospheric MCs \citep{Berger_1996_ApJ,Berger_1998_ApJ,vanBallegooijen_1998_ApJ,Nisenson_2003_ApJ,Chitta_2012_ApJ} is responsible for the wave generation \citep[see the review by][]{Morton_2022_arXiv}. 

Recently, another type of energy transfer attracts attention: magnetic tornado.
Magnetic tornado is a coherent, rotating magnetic field structure passing through from the photosphere to the chromosphere or the base of the corona \citep{Wedemeyer_2012_Nature,Wedemeyer_2013_JPhCS,Wedemeyer_2014_PASJ,Tziotziou_2018_AA}.
The origin of magnetic tornado is the vortex motion of a magnetic element on the photosphere, which is ubiquitously found both observationally \citep{Bonet_2008_ApJ,Bonet_2010_ApJ,Balmaceda_2010_AA} and numerically \citep{Moll_2012_AA,Shelyag_2011_AnGeo,Shelyag_2013_ApJ,Silva_2020_ApJ,Silva_2021_ApJ}.
It should be noted that some chromospheric vortices \citep{Wedemeyer_2009_AA,Morton_2013_ApJ,Park_2016_AA,Liu_2019_NatCo,Shetye_2019_ApJ,Murabito_2020_AA} could be an observational signature of magnetic tornadoes. Therefore, while the observational signature of magnetic tornadoes in the corona may be faint, they may be more frequently occurring phenomena than currently observed. For more information about vortices in the solar atmosphere, please refer to the review by \citet{Tziotziou_2023_SSRv}. A recent numerical simulation by \citet{Wedemeyer_2012_Nature} demonstrated that magnetic tornadoes are capable of transporting sufficient energy to heat the quiet Sun corona. Additionally, the merger of MCs can trigger magnetic tornadoes, which are efficient energy carriers \citep{Finley_2022_AA}. In light of the fact that the merger of MCs frequently occurs on the photosphere \citep{Berger_1996_ApJ,Berger_1998_ApJ,Keys_2011_ApJ,Iida_2012_ApJ}, magnetic tornadoes may have a vital role in coronal heating.

Despite its recognized significance, the underlying physics responsible for the efficient energy transfer mechanisms in magnetic tornadoes remain a subject of investigation. In this study, we address this remained issue through the use of numerical simulations. The rest of this paper is organized as follows. Section \ref{sec:setup} provides the numerical method used in this work. The numerical results and its analyses are shown in Section \ref{sec:results}. The possible implications to the broader context are discussed in Section \ref{sec:discussion}.

\vspace{1em}
\section{Numerical model} 
\label{sec:setup}

We perform a three-dimensional numerical simulation that seamlessly covers the upper part of the solar convection zone and the corona. To this end, we use RAMENS\footnote{RAdiation Magnetohydrodynamics Extensive Numerical Solver} code, in which we solve the compressible magnetohydrodynamic equations with gravity, radiation, and thermal conduction. The basic equations are given as follows.
\begin{align}
\label{eq:mhd_eqs}
    & \frac{\partial \rho} {\partial t} + \nabla \cdot (\rho \boldsymbol{v} )  = 0, \\
    & \frac{\partial (\rho \boldsymbol{v})}{\partial t} 
      +\nabla \cdot \left[ \rho \boldsymbol{v} \boldsymbol{v}
      + \left( p+\frac{\boldsymbol{B}^2}{8\pi} \right) \boldsymbol{ \underbar I}
      - \frac{\boldsymbol{B} \boldsymbol{B}}{4 \pi}  \right]
      = \rho \boldsymbol{g}, \\
    & \frac{\partial \boldsymbol{B}}{\partial t}+\nabla \cdot (\boldsymbol{vB}-\boldsymbol{Bv})=0, \\
    & \frac{\partial e}{\partial t} 
      + \nabla \cdot \left[
     \left(e+p+\frac{\boldsymbol{B}^2}{8\pi}\right)\boldsymbol{v} -\frac{1}{4\pi} \boldsymbol{B}(\boldsymbol{v}\cdot\boldsymbol{B}) \right] \\
    & =\rho \boldsymbol{g}\cdot \boldsymbol{v}+Q_{\mathrm{cnd}}+Q_{\mathrm{rad}} \nonumber , 
\end{align}
where $\rho$ is the mass density,  $\boldsymbol{v}$ is the gas velocity, $\boldsymbol{B}$ is the magnetic field, $e=e_{\rm int} + \rho \boldsymbol{v}^2/2 + \boldsymbol{B}^2/8\pi$ is the total energy density, $e_\mathrm{int}$ is the internal energy density,
$p$ is the gas pressure, $\boldsymbol{g}$ is the gravitational acceleration, and $\boldsymbol{\underbar I}$ is unit tensor.
$Q_\mathrm{cnd}$ and $Q_\mathrm{rad}$ denote the heating by thermal conduction and radiation, respectively. 

The radiation $Q_{\rm rad}$ is given by a combination of optically thick and thin components. In calculating the optically-thick radiation, the radiative transfer is solved under the local thermodynamic equilibrium (LTE) approximation. For simplicity, the gray approximation is applied in the convection zone. The optically thin radiation is calculated from the loss function retrieved from the CHIANTI atomic database ver. 7.1 \citep{Dere_1997_AAS,Landi_2012_ApJ}. The loss function is extrapolated to the lower-temperature range following \citet{Goodman_2012_ApJ}. The equation of state is computed based on the LTE assumption, considering the six most abundant elements in the solar atmosphere (H, He, C, N, O, Ne). The field-aligned thermal conduction of a fully-ionized plasma \citep{Spitzer_1953_PhysRev} is employed to calculate $Q_{\rm cnd}$. Although the assumption of full ionization is invalid in the chromosphere, it does not significantly influence the simulation because the thermal conduction in the chromosphere is minor. The detailed numerical procedure is found in \citet{Iijima_2016_PhD}.

We consider the loop-aligned simulation domain that extends from the upper convection zone to the top of the coronal loop, \added{corresponding to one half of a symmetric closed loop}. For simplicity, we ignore the curvature (non-vertical nature) of the coronal loop. Letting $x$- and $y$-axes be horizontal and $z$-axis be vertical, the size of the simulation domain is set to $3 \mathrm{\ Mm} \times 3 \mathrm{\ Mm} \times \ 15 \mathrm{\ Mm}$ in the $(x, y, z)$ directions. The grid size is uniformly set to 25 km in $x$ and $y$ directions and 50 km in $z$ direction. The periodic boundary conditions are applied in $x$ and $y$ directions.
The bottom boundary condition is \HI{}{open for flow, mimicking the convective energy transport from deep convection zone} (see \citet{Iijima_2016_PhD} for detail).
\replaced{At the top boundary (top of the coronal loop), we impose the reflective boundary, across which $v_z$, $B_x$, and $B_y$ are antisymmetric and the other variables are symmetric}{To ensure complete reflection of the Poynting flux at the top boundary (top of the coronal loop), we apply a reflective boundary that sets $v_z$, $B_x$, and $B_y$ to zero, while the other variables take on the same values as those one grid below. The reflected Poynting flux corresponds to the one injected from the other side of the loop.}
To sustain the million-kelvin corona, we impose an artificial heating at the top boundary so that the temperature $T$ is fixed to $10^6$ K at the top. This artificial heating does not violate the scope of this work because our interest is in the energy transport, not the amount of heating.

The initial ($t=0$) condition in the convection zone is given by Model S \citep{Christensen-Dalsgaard_1996_Science}. Above the surface, the initial condition is calculated by the isothermal stratification. After $5 \ \mathrm{hour}$ of integration, the convection is relaxed to a quasi-steady state, in which the enthalpy flux injected from the bottom boundary nearly equals the radiative flux. Then, we impose a uniform vertical magnetic field of $10 \ \mathrm{G}$, and integrate another $1.5 \ \mathrm{hour}$. We analyze the numerical data in the last 1 hour of the simulation: $19800 \ \mathrm{s} \leq t \leq 23400 \ \mathrm{s}$.

\section{Results} 
\label{sec:results}

\subsection{Simulation overview}
\label{sec:overview}

In this study, we focus on the transient energy transfer from the photosphere to the corona via magnetic tornado. 
To quantitatively discuss the physical properties of magnetic tornado, the background atmosphere as well as the tornado itself is preferred to be as realistic as possible.
In this section, we first overview the background stratification of the model atmosphere to investigate its reality.
For better visualization, we define the $xy$-averaged variables of $F$ as follows.
\begin{align}
    \langle F \rangle_{xy} (z,t) = \frac{1}{L^2} \int_0^L \int_0^L F (x,y,z,t) dxdy,
    \label{eq:xy_average}
\end{align}
where $[0,L] \times [0,L]$ represents the simulation domain in the $xy$ plane. We note that $L = 3 {\rm \ Mm}$ in our numerical setup.

\begin{figure}[!t]
  \centering
  \includegraphics[width = 8cm]{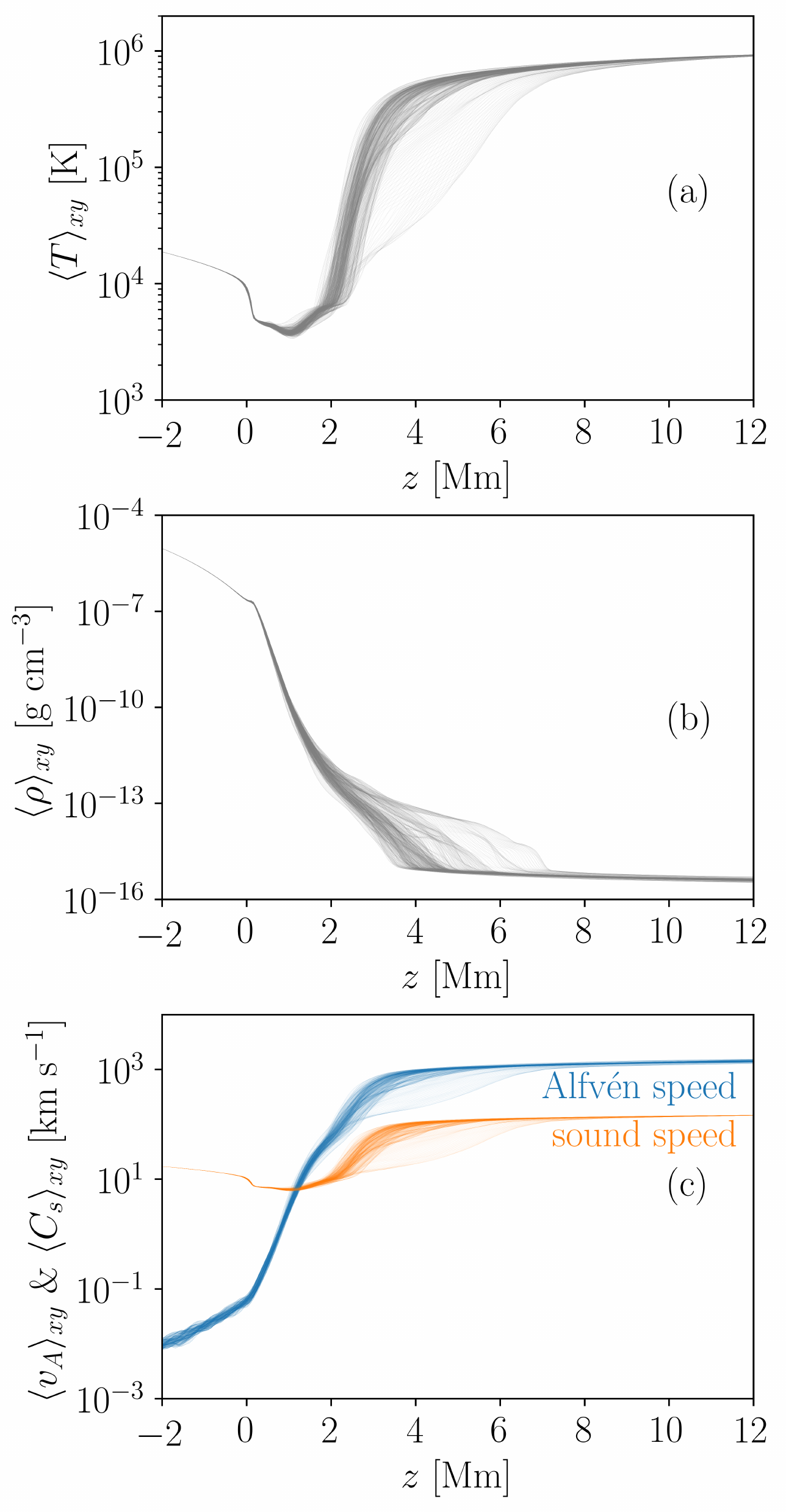}
  \caption{The probability distributions of the horizontally averaged (a) temperature $\langle T \rangle_{xy}$,  (b) density $\langle \rho \rangle_{xy}$, (c) sound  and Alfv{\'e}n speeds  $\langle v_{A} \rangle_{xy}$ with respect to time $t$.}
  \label{fig:pdf}
\end{figure}

\begin{figure*}[!t] 
  \centering
  \includegraphics[width = 18cm]{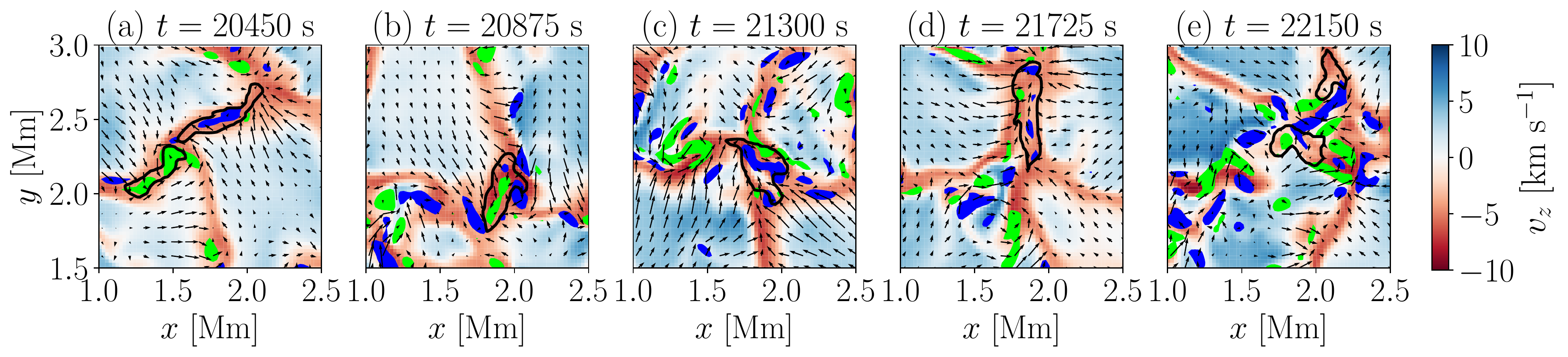}
  \caption{Time sequence of the photospheric magneto-convection at $z=0.0$ Mm. 
  The black contour represents the boundary of MCs, which is defined by $B_z>250$ G. The red-blue color map shows the vertical velocity ($v_z$) and the black arrows illustrate the horizontal velocity field ($\boldsymbol{v}_t=(v_x,v_y)$). Areas filled with blue and green are where the vorticity is strong (green: $\omega_z > 0.05 \ \rm s^{-1}$, blue: $\omega_z < -0.05 \ \rm s^{-1}$. \replaced{(An animation of this figure is available in the online journal.)}{An animation of this figure is available that displays the MC merging event in a period of $1700$ s.}}
  \label{fig:bzwzvtvz}
  \vspace{2em}
\end{figure*}

Figure \ref{fig:pdf} shows the probability distributions of $\langle T \rangle_{xy}$, $\langle \rho \rangle_{xy}$, $\langle v_{A} \rangle_{xy}$, and $\langle C_s \rangle_{xy}$ with respect to $t$, where $v_{A}$ and $C_s$ are Alfv\'en and sound speeds, respectively, which are given by
\begin{align}
    v_{A} = \frac{B_z}{\sqrt{4 \pi \rho}}, \hspace{2em} C_s = \sqrt{\frac{\gamma p}{\rho}}.
\end{align}
The top panel of Figure \ref{fig:pdf} shows that our simulation reproduces 1. the observed solar surface ($z=0.0$ Mm) temperature ($\langle T \rangle_{xy} \approx 5800 {\rm \ K}$), 2. temperature minimum at $z \approx 1.0 {\rm \ Mm}$, and 3. transition region formed at $z \approx 2.5 {\rm \ Mm}$.
These properties are consistent with the solar atmosphere reconstructed from observations \citep{Vernazza_1981_ApJS,Fontenla_1993_ApJ,Avrett_2008_ApJS}.
The coronal Alfv\'en speed ($\approx 1000 {\rm \ km \ s^{-1}}$) is also consistent with observation \citep{Tomczyk_2007_Science,Tomczyk_2009_ApJ}, indicating that the imposed magnetic field has a realistic magnitude.
Moreover, the equipartition layer (where $v_A \approx C_s$) is found in the chromosphere ($z \approx 1.0 \ \rm Mm$), where two types of MHD waves (fast and slow magnetoacoustic waves) couple with each other \citep[i.e., mode conversion, ][]{Schunker_2006_MNRAS,Khomenko_2012_ApJ,Shoda_2018_ApJ,Wang_2021_ApJ}. In light of the fact that the mode conversion likely takes place in the chromosphere \citep{Kontogiannis_2010_AA,Stangalini_2011_AA}, the existence of the equipartition layer in the chromosphere is another possible feature that validates our model.

In summary, our model is capable of reproducing the solar atmosphere that is compatible with observations. This agreement validates the reality of physical processes occurring in the simulation, which are analyzed in the following sections.

\subsection{Onset of Magnetic Tornado}
\label{sec:merger-of-mcs}

Since the plasma beta is higher than unity on the photosphere, magnetic field is passively advected by the convective motion, forming the localized magnetic patches in the intergranular lanes, i.e., MCs.
Generation of MHD waves on the photosphere is attributed to the dynamics of MCs, and thus, the physical properties of the MCs are crucial in understanding the energy transfer from the photosphere to the corona.

Figure \ref{fig:bzwzvtvz} shows the time evolution of MCs (defined by where $B_z > 250 \ \rm G$) on the photosphere ($z=0.0 \ \rm Mm$), with the red-blue color map representing the vertical velocity field. Areas filled in green and blue are where the vertical component of vorticity $\omega_z=(\nabla \times \boldsymbol{v})_z$ is large. Two isolated MCs (Panel (a)) are passively concentrated and merged by the converging granular motion (Panel (b)). 
The merged MC experiences a continuous deformation (Panels (c) and (d)) until it splits into two (Panel (e)). 

\begin{figure}[!t] 
  \centering
  \includegraphics[width = 8cm]{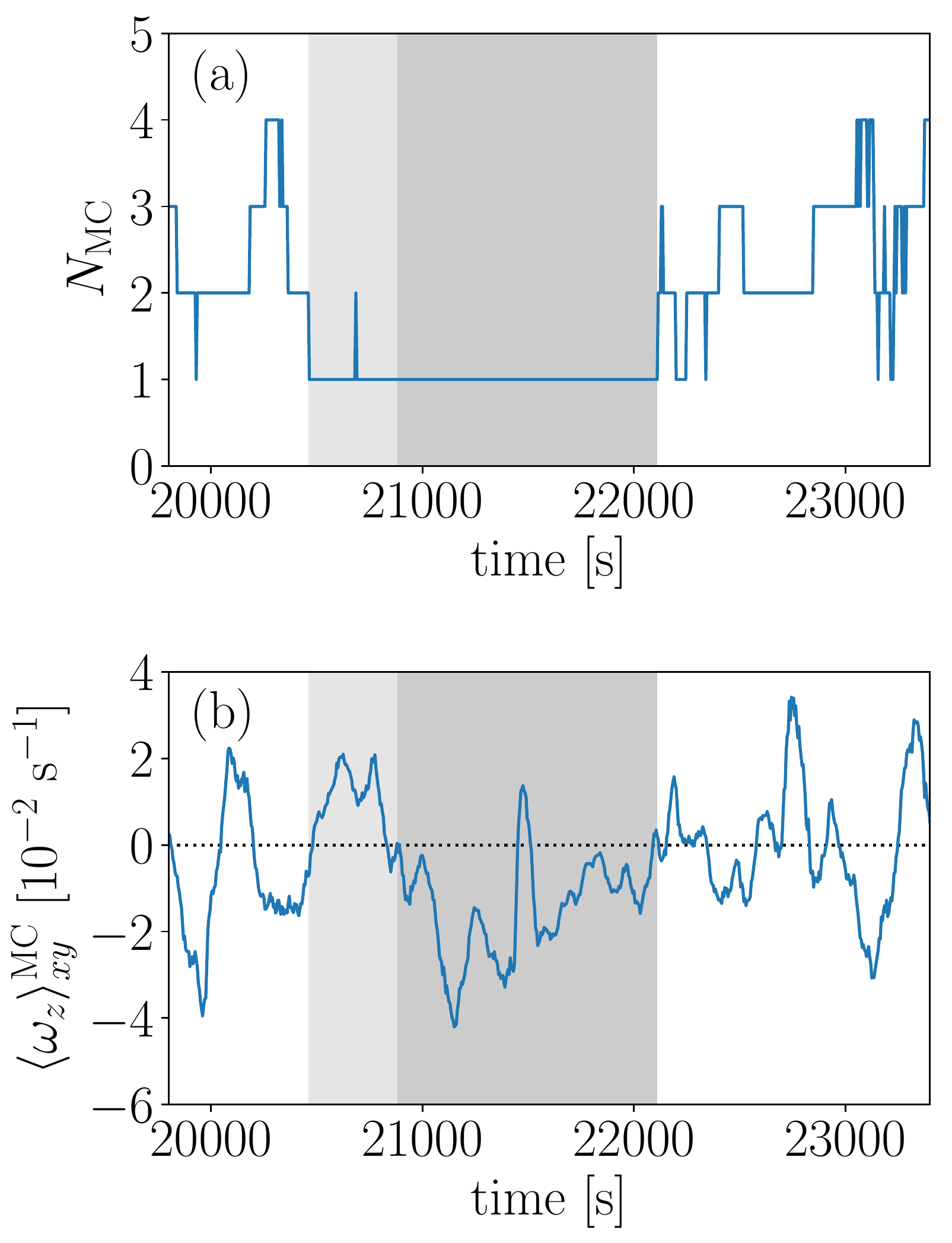}
  \caption{Time evolution of (a) the number of MCs $N_{\rm MC}$ and (b) the mean vertical vorticity over MCs $\langle \omega_z \rangle_{xy}^{\rm MC}$. The gray transparent regions correspond to the time invervals of Phase-1 (brighter part) and Phase-2 (darker part).}
  \label{fig:num_mc}
  \vspace{2em}
\end{figure}

To quantitatively define the period of the MC merging (called hereafter the ``MC merger period''), we show in Figure~\ref{fig:num_mc}a the time evolution of the number of MCs ($N_{\rm MC}$), which is defined by the number of the connected areas of $B_z > 250 \ \rm G$, whose areas are larger than $22500 \ \rm km^2$.
$N_{\rm MC}$ is kept nearly maintained to unity in $20460 \ {\rm s} \leq t \leq 22110 \ {\rm s}$, and thus, this period is defined as the MC merger period.
It should be mentioned that we have excluded the case where $N_{\rm MC} = 2$ at $t=20700 \ \rm s$ because the two magnetic concentrations are separated by only one numerical pixel at this time.

The vertical vorticity $\omega_z$ shown in Figure \ref{fig:bzwzvtvz} indicates that the vorticity on the merged MC changes its sign during the MC merger period.
This effect is more quantitatively seen by defining the mean vorticity of MC as follows,

\begin{align}
\langle \omega_z \rangle^{\rm MC}_{xy} =\frac{\displaystyle \int_{\rm MC} \omega_z \ dxdy }{\displaystyle \int_{\rm MC} dxdy },
\end{align}
where the subscript ``MC" refers to the area covered by MCs in the whole $z=0.0$ Mm plane. Figure~\ref{fig:num_mc}b shows the time evolution of $\langle \omega_z \rangle^{\rm MC}_{xy}$.
The sign of $\langle \omega_z \rangle^{\rm MC}_{xy}$ changes during the MC merger period at $t=20880 \ \rm s$. Since the different sign in $\langle \omega_z \rangle^{\rm MC}_{xy}$ should result in the different magnetic-field structure in the upper atmosphere, we shall divide the MC merger period into two phases: $20460 \ {\rm s} \leq t < 20880 \ {\rm s}$ (called ``Phase-1") and $20880 \ {\rm s} \leq t \leq 22110 \ {\rm s}$ (called ``Phase-2").

\begin{figure}[!t] 
  \centering
  \includegraphics[width = 8cm]{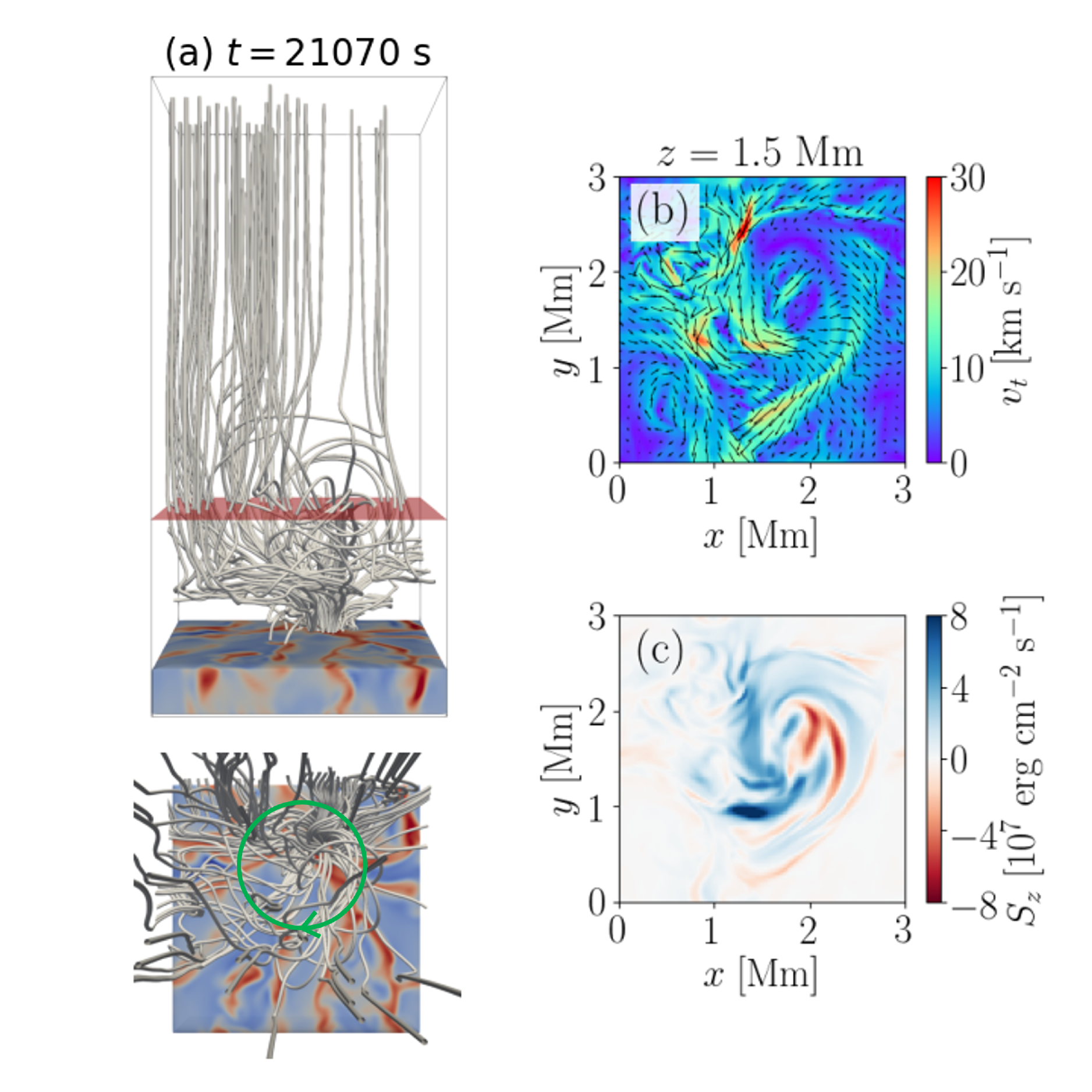}
  \caption{Several physical quantities in the presence of the magnetic tornado triggered by the photospheric vortices in Phase-1. 
(a): The magnetic field lines rooted in the merged MC are shown, along with the vertical velocity maps from $z=-1.0$ Mm to $z=0.0$ Mm in red-blue color. The red plane represents the chromospheric layer ($z=1.5$ Mm), and the green lines indicate the horizontal twisting direction of the magnetic field lines. The upper panel provides a view along $y$-direction from a reference point at $y=0.0$ Mm, while the lower panel provides a view along $z$-direction from a reference point at $z=6.0$ Mm.
(b): The horizontal velocity field $\boldsymbol{v}_t$ is shown in black arrows, and its amplitude is indicated by the color at $z=1.5$ Mm. 
(c): The vertical Poynting flux $S_z$ at $z=1.5$ Mm. 
Each panel is taken at $t=21070$ s. 
We note that the time lag between the formation of the positive $\langle \omega_z \rangle^{\rm MC}_{xy}$ in Phase-1 and the appearance of the chromospheric swirling pattern of $\boldsymbol{v}_t$ and $S_z$ in the panels reflects the propagation of the vortex through the magnetic field from $z=0.0$ Mm to $z=1.5$ Mm. \replaced{(An animation of this figure is available in the online journal.)}{An animation of this figure is available that shows the temporal evolution of the magnetic field lines, $\boldsymbol{v}_t$ and $S_z$ of the magnetic tornado in a period of $345$ s.}}
  \label{fig:phase-1}
  \vspace{2em}
\end{figure}

\begin{figure}[!t] 
  \centering
  \includegraphics[width = 8cm]{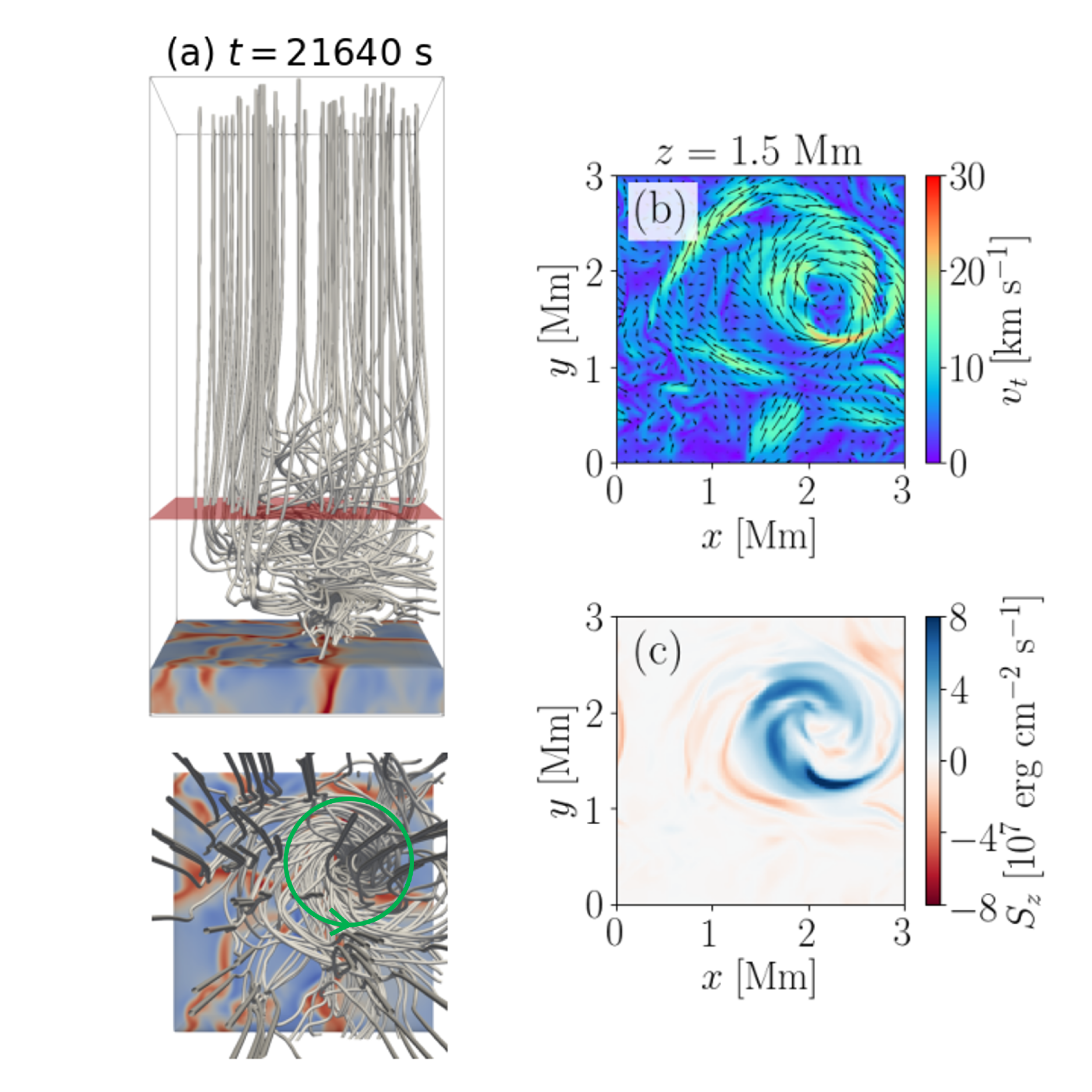}
  \caption{Same as Figure \ref{fig:phase-1}, but each panel is taken at $t=21640$ s, which corresponds to Phase-2. \replaced{(An animation of this figure is available in the online journal.)}{An animation of this figure is available that shows the temporal evolution of the magnetic field lines, $\boldsymbol{v}_t$ and $S_z$ of the magnetic tornado in a period of $470$ s.}}
  \label{fig:phase-2}
  \vspace{2em}
\end{figure}

Figure \ref{fig:phase-1} and Figure \ref{fig:phase-2} display the three-dimensional magnetic field lines alongside the chromospheric ($z=1.5$ Mm) horizontal velocity $\boldsymbol{v}_t=(v_x.v_y)$, facilitating a comprehensive analysis of the possibility of a magnetic tornado formation on the merged magnetic configuration. 
The magnetic field lines shown in Figure \ref{fig:phase-1} (Figure \ref{fig:phase-2}) are coherently twisted in a clockwise (counerclockwise) direction from the corona down into the photosphere, whilst inducing a swirling pattern of the chromospheric horizontal velocity in the counterclockwise (clockwise) direction.
Our results suggest that the former (latter) magnetic and velocity field patterns are triggered by the photospheric vortices present in Phase-1 (Phase-2).
In response to the sign reversal in $\langle \omega_z \rangle^{\rm MC}_{xy}$ observed during the transition from Phase-1 to Phase-2, the magnetic tornado undergoes a change in its rotational direction from counterclockwise to clockwise.

Figure \ref{fig:phase-1} and Figure \ref{fig:phase-2} also provide insight into the vertical Poynting flux $S_z=-[(\boldsymbol{v} \times \boldsymbol{B}) \times \boldsymbol{B}]_z / 4\pi$ at the chromospheric height ($z=1.5$ Mm). 
The distinctive ring-like patterns observed in the chromospheric $S_z$ roughly track the twisted magnetic fields, providing compelling evidence that the magnetic tornado serves as an energy conduit through the solar atmosphere.
The diameter of the ring-like patterns is approximately 1.5-2.0 Mm, which is consistent with the typical size of the chromospheric swirls (approximately 2.0 Mm) reported in previous studies \citep{Wedemeyer_2012_Nature,Shetye_2019_ApJ}.
Such large-scale ring-like patterns ($>1.5$ Mm) are produced only when the magnetic tornado rooted in the merged MC exists.

\added{The twisted magnetic field of the magnetic tornado appears to be confined below the bottom of the corona ($< 2$-$4$ Mm), due to two key factors. Firstly, the nonlinearity of transverse Alfv{\'e}n waves propagating along $z$-direction, which is indicated by $v_t$ normalized by $v_A$, is significantly higher in the chromosphere (i.e., $v_t / v_A>1$) than in the corona (i.e., $v_t / v_A<0.1$). As a result, the twisting is more prominent in the chromosphere. 
Secondly, Alfv{\'e}n waves in the corona propagate much faster than in the chromosphere due to the difference in local Alfv{\'e}n speed (typically, $10 \rm km s^{-1}$ in the chromosphere and $1000 \rm km s^{-1}$ in the corona, see Figure \ref{fig:pdf}c). Consequently, the release of local twisting of the magnetic field is much more rapid in the corona.
Regarding our primary focus of investigating the energy transfer mechanism from the lower atmosphere to the corona, it is worth noting that the reflected Alfv{\'e}n waves from the upper boundary have a minimal impact. This is due to the fact that the energy of the reflected Alfv{\'e}n waves from the top boundary is much lower than that of the chromospheric Alfv{\'e}n waves, typically below $10\%$. As a result, the feedback from the upper boundary does not significantly alter the energy transfer process.}

\subsection{Energy transfer by magnetic tornado}
\label{sec:poynting}

To investigate the amount and efficiency of energy transfer by the magnetic tornado, we investigate the variation in the Poynting flux associated with the magnetic tornado.
For this purpose, we calculate the $z$ component of the Poynting flux given by
\begin{align}
    S_z = S_z^{\rm shear} + S_z^{\rm emerge}, 
\end{align}
where
\begin{align}
    S_z^{\rm shear} = - \frac{B_z}{4 \pi} \left( \boldsymbol{v}_t \cdot \boldsymbol{B}_t \right), \hspace{1em} S_z^{\rm emerge} = v_z \frac{\boldsymbol{B}_t^2}{4 \pi},
\end{align}
where $\boldsymbol{B}_t=(B_x,B_y)$.
$S_z^{\rm shear}$ is associated with the transverse displacement of the vertical magnetic field, while $S_z^{\rm emerge}$ corresponds to the vertical motion of the horizontal magnetic field.
We note that $S_z^{\rm shear}$ and $S_z^{\rm emerge}$ are often interpreted as energy transfers by shearing motion and flux emergence, respectively \citep{Welsch_2015_PASJ,Cranmer_2019_ARAA}.

\begin{figure}[!t] 
  \centering
  \includegraphics[width = 8cm]{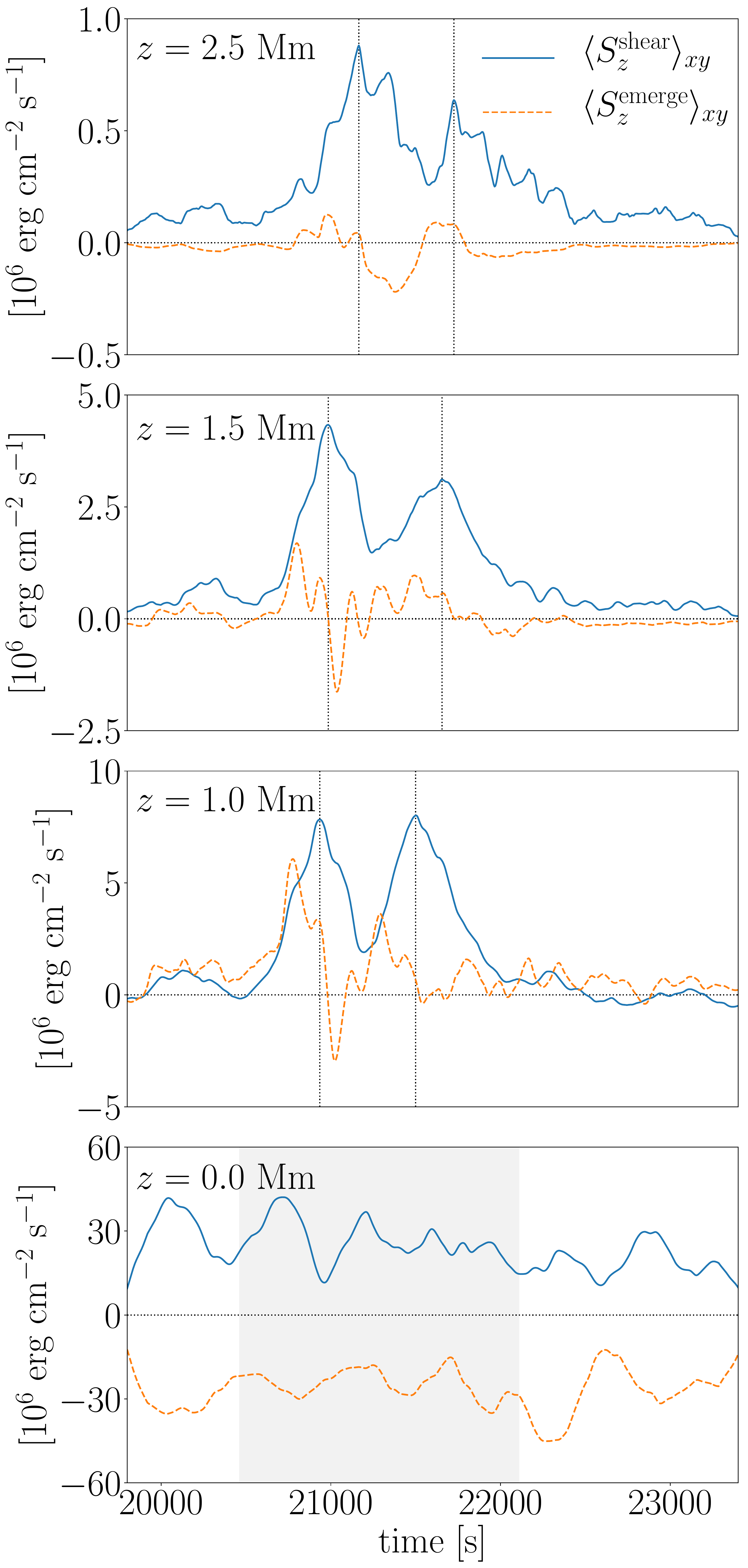}
  \caption{Time series of $\langle S_z^{\rm shear} \rangle_{xy}$ (blue solid line) and $\langle S_z^{\rm emerge} \rangle_{xy}$ (orange dashed line) at a height of $z=0.0$ Mm (photosphere), $z=1.0$ Mm (lower chromosphere), $z=1.5$ Mm  (upper chromosphere), and $z=2.5$ Mm (transition region). The vertical black dotted lines mark the positions of the peaks of $\langle S_z^{\rm shear} \rangle_{xy}$ that correspond to Phase-1 and Phase-2. The gray transparent region in the bottom panel indicates the MC merger period.}
  \label{fig:sztl}
\end{figure}

Figure \ref{fig:sztl} shows the time evolution of $\langle S_z^{\rm shear} \rangle_{xy}$ and $\langle S_z^{\rm emerge} \rangle_{xy}$ (see Eq.~\eqref{eq:xy_average}) for four different heights: $z=0.0$ Mm, $z=1.0$ Mm, $z=1.5$ Mm, and $z=2.5$ Mm.
To enhance the features associated with the magnetic tornado triggered by the MC merger,
we apply the low-pass filter with a cutoff frequency of $4.2$ mHz. This cutoff frequency is chosen to be between the typical correlation time of the shear flows within intergranular lanes $\sim 40$-$100 \ \rm s$ \citep{vanBallegooijen_2011_ApJ,Fedun_2011_AnGeo} and the duration of Phase-1 $420 \ \rm s$.
The following properties are inferred from this analysis.
\begin{enumerate}
\item Phase-1 and Phase-2 correspond to the dual maxima presented in $\langle S_z^{\rm shear} \rangle_{xy}$ at heights of $z=1.0$, $1.5$, and $2.5$ Mm. 
Interestingly, we find no clear signature of Phase-1 nor Phase-2 in $\langle S_z^{\rm shear} \rangle_{xy}$ at $z=0.0$ Mm, indicating that the onset of the magnetic tornado does not enhance the net Poynting flux on the photosphere.
\item In the presence of the magnetic tornado, the time-averaged Poynting flux measured at the transition region ($z=2.5 \ \mathrm{Mm}$) is $4.2 \times 10^5 {\rm \ erg \ cm^2 \ s^{-1}}$, which is four times larger than that without the magnetic tornado ($\approx 1.0 \times 10^5 {\rm \ erg  \ cm^2 \ s^{-1}}$). Magnetic tornado is thus probably a more efficient energy transfer mechanism than the simple convective buffeting of the magnetic field line.
\item The temporal evolution of $\langle S_z^{\rm shear} \rangle_{xy}$ exhibits altitude-dependent characteristics, with the timing of its initial and subsequent maxima varying with height. 
Specifically, the first (second) maximum is observed at $t=20935 s$ ($21500 \ \rm s$) in $z=1.0 \ \rm Mm$, $20985 \ \rm s$ ($21655 \ \rm s$) in $z=1.5 \ \rm Mm$, and $t=21165 \ \rm s$ ($21725 \ \rm s$) in $z=2.5 \ \rm Mm$. This temporal difference corresponds to the energy propagation time of the magnetic tornado to each layer.

\item $S_z^{\rm emerge}$ is smaller than $S_z^{\rm shear}$ for the majority of the time.
In particular, $S_z^{\rm emerge}$ is negative in $z=0.0$ Mm. Thus, in our specific setup, the vertical motion of the horizontal magnetic field plays minor or negative roles in the energy transfer. Although a similar trend is found in the previous work \citep{Finley_2022_AA}, care needs to be taken in drawing a conclusion because the dynamics of flux emergence are significantly affected by the bottom boundary condition \citep{Rempel_2014_ApJ}. We need a more detailed analysis to clarify the role of flux emergence in coronal heating \citep{Wang_2020_ApJ,Wang_2022_SolPhys} using, for example, a numerical model with sufficiently deep convection zone \citep[e.g.,][]{Hotta_2019_SciA}.
\end{enumerate}

\begin{figure}[!t] 
  \centering
  \includegraphics[width = 7cm]{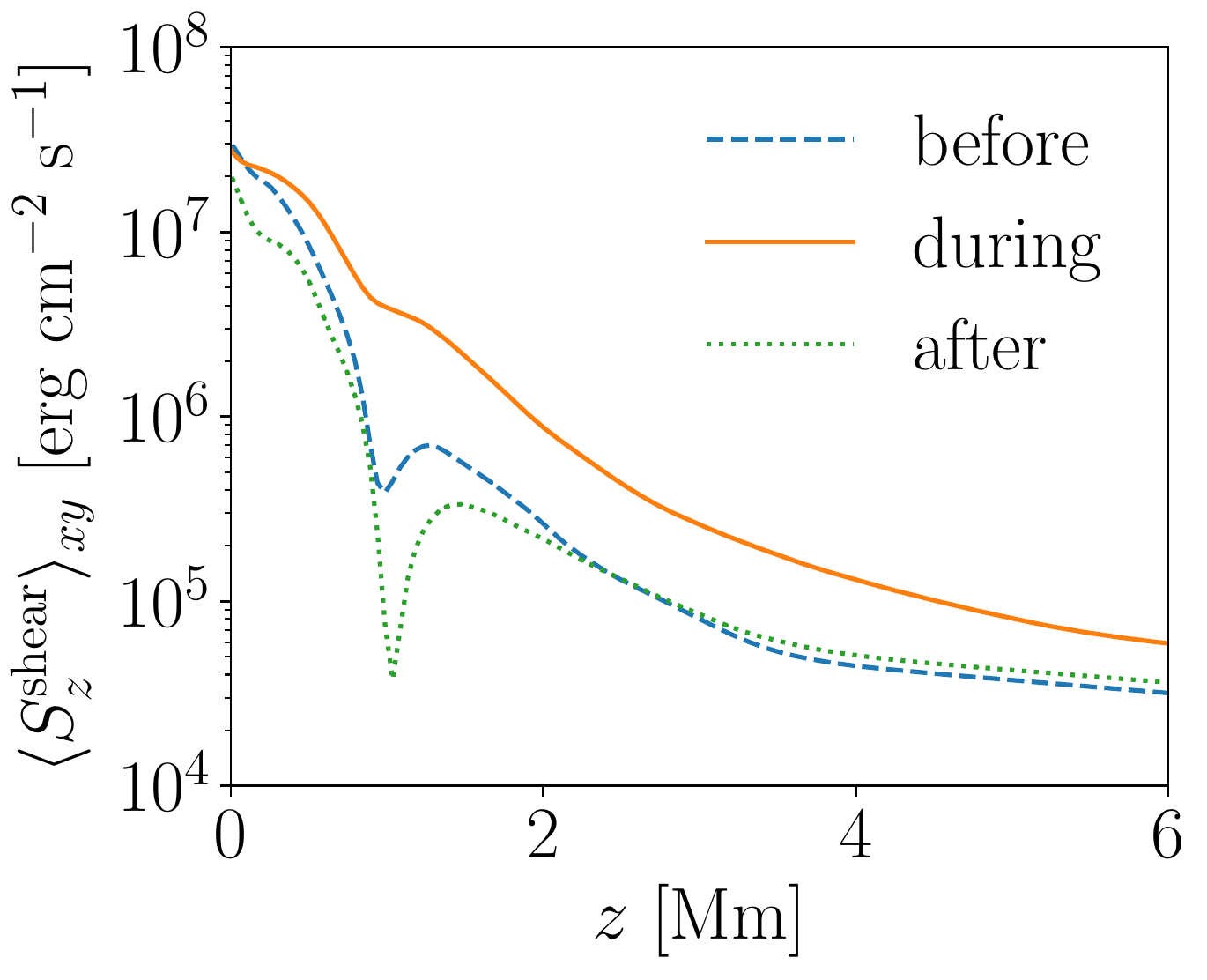}
  \caption{$\langle S_z^{\rm shear} \rangle_{xy}$ versus height $z$ averaged over three time intervals: before ($19800 \ \mathrm{s} \leq t < 20460 \ \mathrm{s}$, blue dashed line), during ($20460 \ \mathrm{s} \leq t \leq 22110 \ \mathrm{s}$, orange solid line), and after ($22110 \ \mathrm{s} < t \leq 23400 \ \mathrm{s}$, green dotted line) tornado.}
  \label{fig:szt-alongz}
\end{figure}

Figure \ref{fig:szt-alongz} illustrates the time-averaged $\langle S_z^{\rm shear} \rangle_{xy}$ as a function of height $z$ for three distinct time intervals: ``before tornado" phase ($19800 \ \mathrm{s} \leq t < 20460 \ \mathrm{s}$), ``during tornado" phase ($20460 \ \mathrm{s} \leq t \leq 22110 \ \mathrm{s}$, identical to the MC merger period), and ``after tornado" phase ($22110 \ \mathrm{s} < t \leq 23400 \ \mathrm{s}$).
$\langle S_z^{\rm shear} \rangle_{xy}$ at $z=0.0 \ \mathrm{Mm}$ remains nearly constant regardless of the presence of the magnetic tornado (as inferred from Figure \ref{fig:sztl}).
Meanwhile, in the presence of the magnetic tornado, $\langle S_z^{\rm shear} \rangle_{xy}$ decreases more gradually in height, especially below the corona ($0.0 \ \rm Mm \leq z \leq 2.5 \ \rm Mm$).

There are several possible reasons for the smaller decrease of $\langle S_z^{\rm shear} \rangle_{xy}$ during tornado, including the smaller energy reflection and smaller energy dissipation.
To distinguish these two factors, we decompose $\langle S_z^{\rm shear} \rangle_{xy}$ as follows.
\begin{align}
    \langle S_z^{\rm shear} \rangle_{xy} = \langle S_z^{\rm shear,+} \rangle_{xy} - \langle S_z^{\rm shear,-} \rangle_{xy}
\end{align}
where
\begin{align}
    S_z^{\rm shear, \pm} 
    = \frac{1}{4} \rho v_A \left( \boldsymbol{v}_t \mp \frac{\boldsymbol{B}_t}{\sqrt{4 \pi \rho}} \right)^2
    = \frac{1}{4} \rho v_A {\boldsymbol{z}_t^\pm}^2.
\end{align}
Under the approximation of incompressible or reduced MHD \citep{Strauss_1976_PhFl}, $S_z^{\rm shear, \pm}$ represents the energy fluxes of upward and downward Alfv\'en waves, respectively \citep{Elsasser_1950_PhRv, Heinemann_1980_JGR}.
Although the Els\"asser variables are no longer the characteristics of Alfv\'en waves in an inhomogeneous or compressional system \citep{Velli_1989_PRL, Hollweg_2007_JGR, Magyar_2019_ApJ}, they are often still useful in decomposing the propagating direction \citep{Marsch_1987_JGR, Magyar_2019_ApJ}.

\begin{figure}[!t] 
  \centering
  \includegraphics[width = 7.5cm]{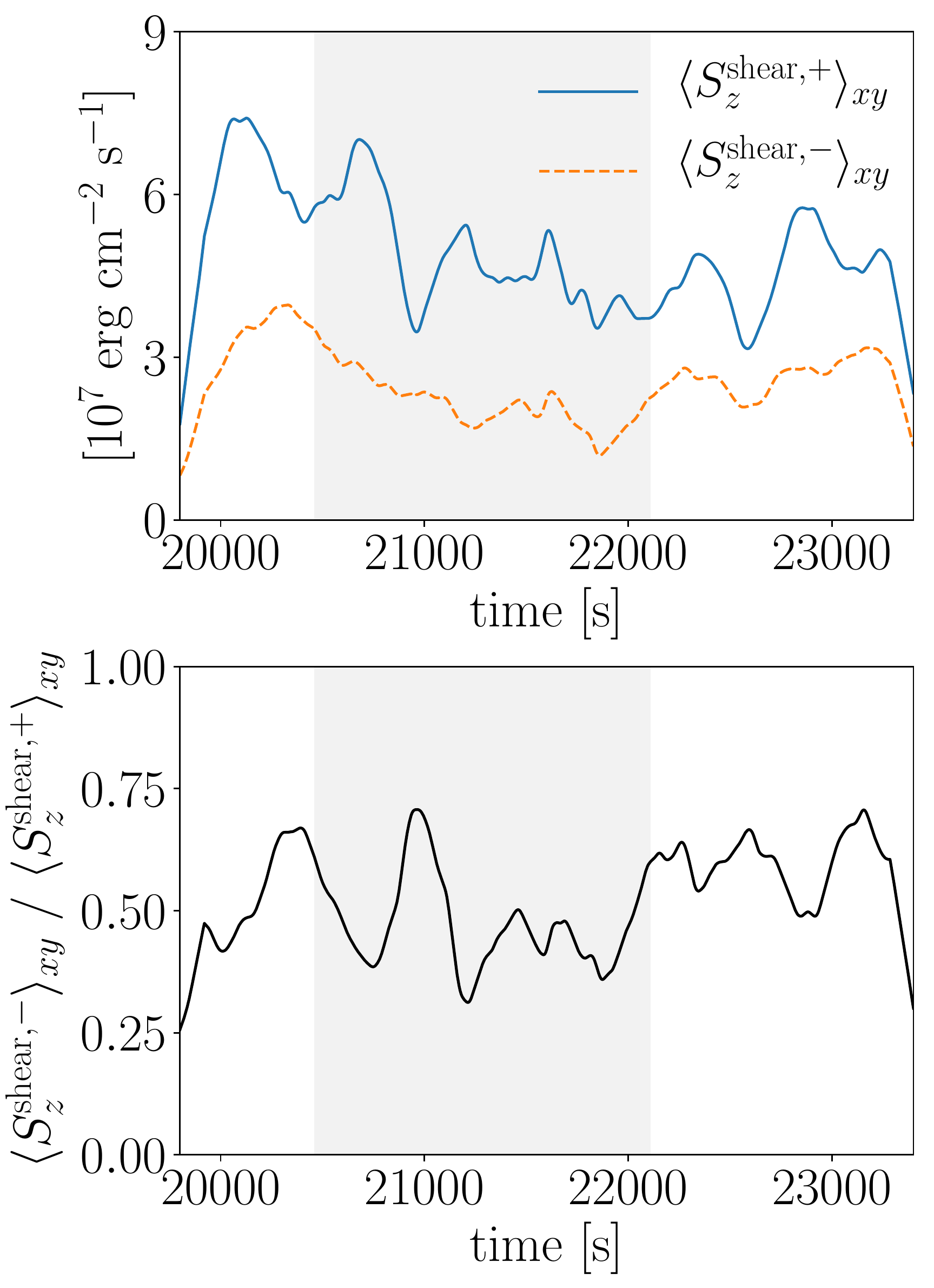}
  \caption{Time series of $\langle S_z^{\rm shear,+} \rangle_{xy}$ (top, blue solid line), $\langle S_z^{\rm shear,-} \rangle_{xy}$ (top, orange dashed line) and the ratio of them $\langle S_z^{\rm shear,+} \rangle_{xy}/\langle S_z^{\rm shear,-} \rangle_{xy}$ (bottom) measured on the photosphere ($z=0.0{\rm \ Mm}$). The gray transparent region shows the MC merger period.}
  \label{fig:reflection}
\end{figure}

\begin{figure*}[!t] 
  \centering
  \includegraphics[width = 15cm]{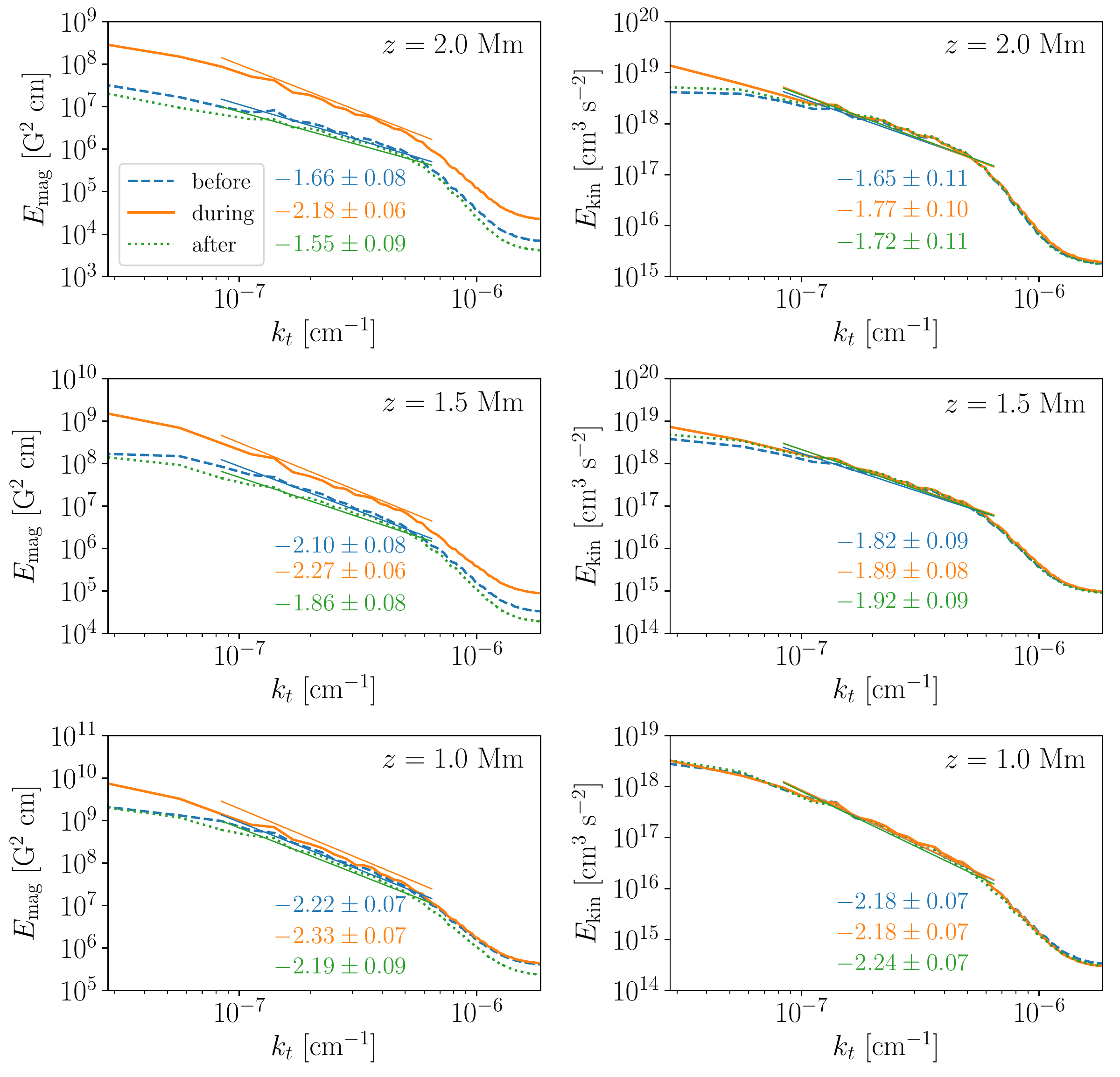}
  \caption{Power spectral densities of transverse magnetic field $E_{\rm mag}$ and transverse velocity $E_{\rm kin}$ for three different heights: $z = 1.0$ (bottom), $1.5$ (middle), and $2.0$ Mm (top).
  Different lines correspond to the time interval of spectrum calculation (blue dashed: before tornado, orange solid: during tornado, green dotted: after tornado). Thin lines show the power-law fittings to the same-colored thick lines in the inertial range, the indices of which are displayed by numbers in each panel.}
  \label{fig:turbulence}
  \vspace{2em}
\end{figure*}

The top panel of Figure \ref{fig:reflection} displays the time series of $\langle S_z^{\rm shear,+} \rangle_{xy}$ and $\langle S_z^{\rm shear,-} \rangle_{xy}$ measured on the photosphere ($z=0.0 \ \mathrm{Mm}$). The bottom panel shows the ratio of the two quantities, $\langle S_z^{\rm shear,-} \rangle_{xy} / \langle S_z^{\rm shear,+} \rangle_{xy}$.
The gray transparent region corresponds to the MC merger period.
We find no significant change in $\langle S_z^{\rm shear,+} \rangle_{xy}$ nor $\langle S_z^{\rm shear,-} \rangle_{xy}$ during tornado.
This means that the magnetic tornado does not affect the energy injection nor reflection.
The smaller decrease in $S_z^{shear}$ along height (Figure \ref{fig:szt-alongz}) is thus (at least partly) attributed to the change in the energy dissipation process.

\subsection{Turbulence during magnetic tornado}
\label{sec:turbulence}

In the preceding section, we demonstrate that the crucial factor responsible for the increased energy transfer to the corona is the reduced dissipation of Poynting flux beneath the corona, rather than the increased injection or reduced reflection.
To further investigate its physical origin, we focus on the efficiency of energy cascading in the presence and absence of the magnetic tornado. 
For this purpose, we define the one-dimensional (reduced) energy spectra of velocity and magnetic field with respect to transverse wave number as follows.
\begin{align}
    &E_{\rm kin} (k_t) = \nonumber \\
    & \frac{1}{\Delta k_t} \int_{k_t \le \sqrt{k_x^2+k_y^2} < k_t + \Delta k_t} dk_x dk_y \left( \frac{L}{2 \pi} \right)^2 \left| \boldsymbol{v}_t \left( k_x, k_y \right) \right|^2,
\end{align}
\begin{align}
    &E_{\rm mag} (k_t) = \nonumber \\
    &\frac{1}{\Delta k_t} \int_{k_t \le \sqrt{k_x^2+k_y^2} < k_t + \Delta k_t} dk_x dk_y \left( \frac{L}{2 \pi} \right)^2 \left| \boldsymbol{B}_t \left( k_x, k_y \right) \right|^2,
\end{align}
where $(k_x, k_y)$ represents the wavenumber in $x$- and $y$-directions respectively, $\Delta k_t$ is a sufficiently small value, and
\begin{align}
    \boldsymbol{v}_t \left( k_x, k_y \right) = \frac{1}{L^2} \int_{[0,L] \times [0,L]} dxdy \ \boldsymbol{v}_t (x,y) \ e^{-i (k_x x + k_y y)},
\end{align}
\begin{align}
    \boldsymbol{B}_t \left( k_x, k_y \right) = \frac{1}{L^2} \int_{[0,L] \times [0,L]} dxdy \ \boldsymbol{B}_t (x,y) \ e^{-i (k_x x + k_y y)}.
\end{align}
It follows, from Parseval's identity, that $E_{\rm kin} (k_t)$ and $E_{\rm mag} (k_t)$ satisfy
\begin{align}
    \int dk_t \ E_{\rm kin} (k_t) &= \langle \boldsymbol{v}_t (x,y)^2 \rangle_{xy},
\end{align}
\begin{align}
    \int dk_t \ E_{\rm mag} (k_t) &= \langle \boldsymbol{B}_t (x,y)^2 \rangle_{xy}.
\end{align}

\begin{figure}[!t] 
  \centering
  \includegraphics[width = 8cm]{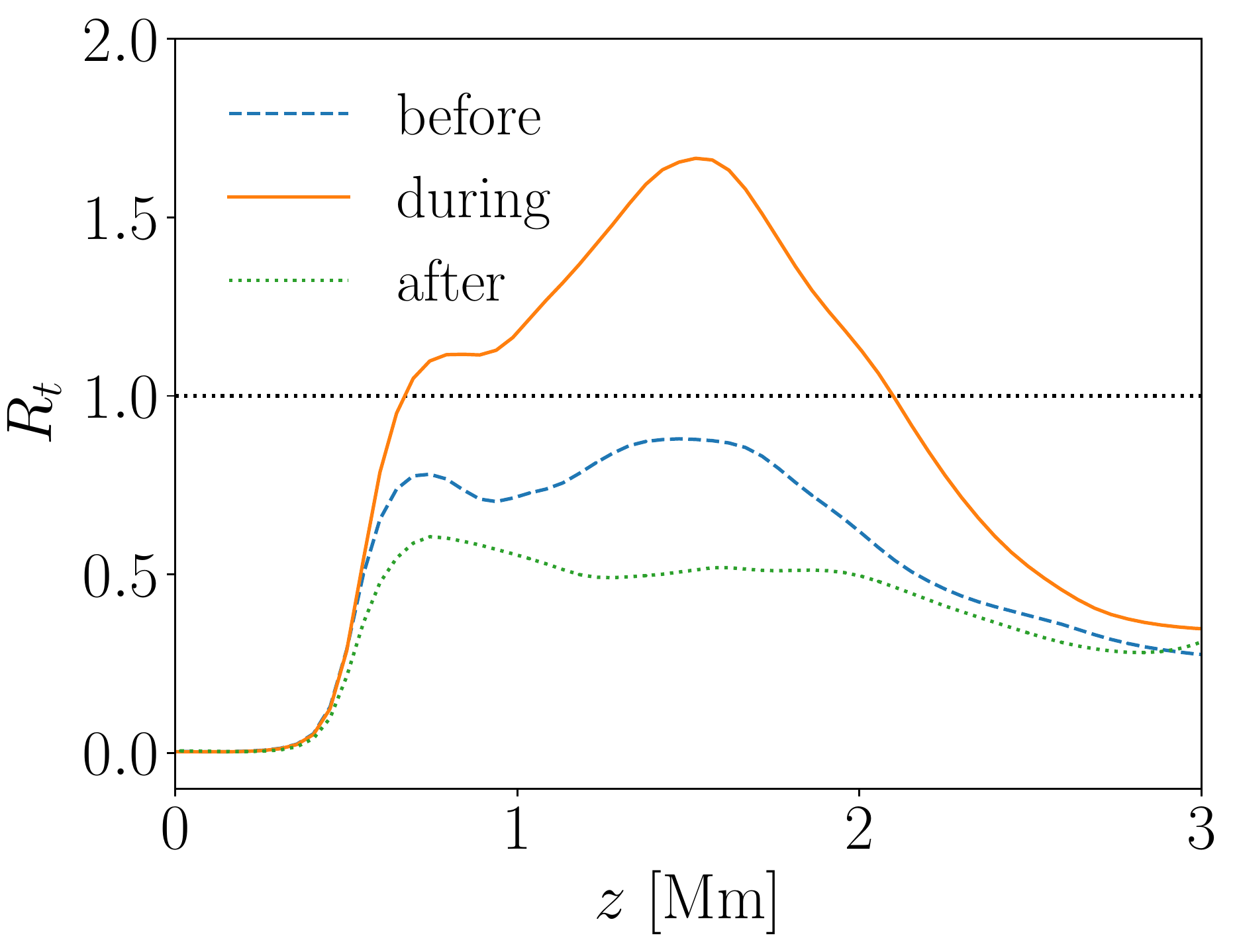}
  \caption{The ratio of the $xy$-averaged transverse magnetic and kinetic energies ($R_t$, see Eq.~\eqref{eq:energy_ratio}), versus height averaged over three time intervals: before ($19800 \ \mathrm{s} \leq t < 20460 \ \mathrm{s}$, blue dashed line), during ($20460 \ \mathrm{s} \leq t \leq 22110 \ \mathrm{s}$, orange solid line), and after ($22110 \ \mathrm{s} < t \leq 23400 \ \mathrm{s}$, green dotted line) the onset of the magnetic tornado.}
  \label{fig:emkt_ratio}
\end{figure}

Figure~\ref{fig:turbulence} shows $E_{\rm kin} (k_t)$ and $E_{\rm mag} (k_t)$ for three different heights ($z = 1.0$, $1.5$, and $2.0$ Mm) averaged over three characteristic time intervals: before, during, and after tornado.
Both $E_{\rm kin} (k_t)$ and $E_{\rm mag} (k_t)$ become flatter as $z$ increases, implying that the turbulence evolves in height.
We find no clear difference in $E_{\rm kin} (k_t)$ for the three time intervals, which means that the presence of the magnetic tornado hardly affects the cascading of kinetic energy.
In contrast, the magnetic energy spectra $E_{\rm mag} (k_t)$ exhibit notable modifications due to the magnetic tornado; $E_{\rm mag} (k_t)$ becomes steeper, in particular at a height of $z=2.0$ Mm.
The increased steepness of the magnetic spectrum implies that the magnetic-energy cascading is weakened in the presence of the magnetic tornado.

Figure \ref{fig:emkt_ratio} shows the $xy$-averaged magnetic-to-kinetic energy ratio for the transverse components, that is,
\begin{align}
    R_t = \frac{\langle \boldsymbol{B}_t^2 / (8 \pi) \rangle_{xy}}{\langle \rho \boldsymbol{v}_t^2 /2 \rangle_{xy}}, \label{eq:energy_ratio}
\end{align}
for the three time intervals.
In the absence of magnetic tornado, the transverse fluctuations are dominated by velocity ($R_t<1$), which is in agreement with the Alfv\'en-wave propagation models in the chromosphere \citep{Cranmer_2005_ApJ,Verdini_2007_ApJ}.
During the existence of magnetic tornado, on the other hand, the transverse fluctuations are mostly magnetic ($R_t>1$) in the chromosphere ($1.0 \ \mathrm{Mm} \le z \le 2.0 \ \mathrm{Mm}$).
The loss of Poynting flux in the chromosphere should thus be affected by the cascading of $E_{\rm mag}$ rather than $E_{\rm kin}$ in the presence of the magnetic tornado.

To summarize, the smaller loss of Poynting flux in the presence of the magnetic tornado is possibly attributed to the change in the magnetic-energy spectrum, or equivalently, magnetic-energy cascading.

\begin{figure*}[!t] 
  \centering
  \includegraphics[width = 13.5cm]{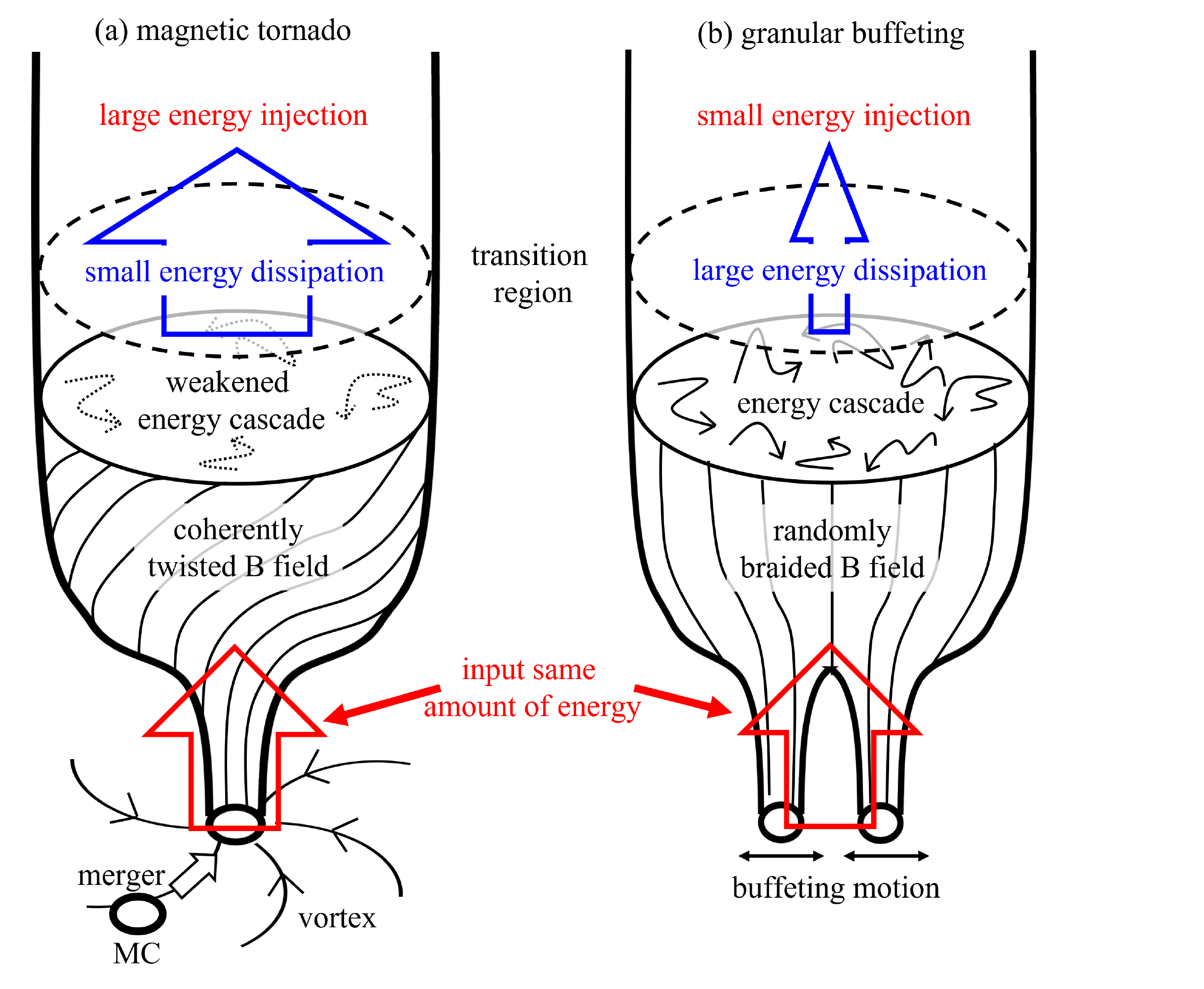}
  \caption{A schematic illustration of the physical processes underlying our findings, with a focus on the contrast between the energy generation and propagation mechanisms in the presence (panel (a)) and absence (panel (b)) of the magnetic tornado.}
  \label{fig:discussion}
  \vspace{2em}
\end{figure*}

\vspace{1em}
\section{Discussion}
\label{sec:discussion}

\subsection{Efficient energy transfer by magnetic tornado}
Our simulation reveals that the merger of MCs serves as a trigger for the onset of a magnetic tornado with the large-scale chromospheric swirling patterns (of which diameter is $>1.5$ Mm), which in turn facilitates the efficient transport of energy to the corona. This finding is in line with previous investigations by \citet{Finley_2022_AA}.
In this section, we further develop our discussion to address the underlying physical process for the enhancement of the coronal energy transport when the magnetic tornado is present, as compared to its absence.

The merged MC exhibits a longer lifetime of 1650 s ($20460 \ \rm s < t < 22110 \ \rm s$) in comparison to the average lifetime of individual MCs, which is typically 300-500 s.
This finding is consistent with prior observations demonstrating that the lifetimes of magnetic bright points (MBPs), which serve as observational signatures of MCs, are prolonged during merging events as compared to the lifetimes of individual MBPs \citep{Keys_2011_ApJ}.
Given that MCs can impede local magnetoconvection and prevent the collapse of photospheric vortices \citep{Finley_2022_AA}, it is plausible that the prolonged lifetimes of MCs observed in this study may facilitate the development of long-lived photospheric vortices, which can in turn give rise to magnetic tornadoes featuring the large-scale chromospheric swirls. The detailed analysis of the condition for magnetic tornadoes to form is a line of the future research.

The analysis of Poynting flux reveals that the magnetic tornado carries a significantly larger energy, four times greater than that of the system without the magnetic tornado. This enhancement cannot be attributed to an increase in energy generation on the photosphere, as there is no enhancement in the upward Poynting flux when the magnetic tornado is formed. 
Instead, we find that the different behaviors in the chromosphere, particularly the smaller loss of Poynting flux due to the weakened energy cascading, appear to be the most promising cause of the efficient energy transfer.
This result can be attributed to the inverse cascade process, which facilitates the transfer of magnetic energy to larger scales than the injection scale, where dissipation mechanisms become less effective, leading to an excess of magnetic energy in comparison to kinetic energy.
This phenomenon has been observed in various models utilizing the reduced MHD framework, which is applicable to low-beta and incompressible plasmas, such as the corona \citep{Rappazzo_2013_ApJ,Rappazzo_2019_ApJ}. 
Whereas, care needs to be taken when applying the inverse cascade scenario to our case, as we focus on the chromosphere, where compressibility can not be ignored, unlike the corona. Furthermore, it is worth noting that the magnetic tornado occurs impulsively, persisting for only a few Alfv\'en travel times across the entire loop, whereas the inverse cascade typically requires more than $100$ Alfv\'en travel times to develop.

The overall scenario found in our simulation is schematically summarized in Figure~\ref{fig:discussion}.
The merger of two MCs induces the long-lived vortex motions inside the resulting merged MC.
Such long-lived vortices generate the coherently twisted magnetic field lines from the photosphere to the corona accompanied by swirling motions: the magnetic tornado.
With the magnetic tornado present, energy cascading in the chromosphere is weakened, leading to enhanced energy injection to the corona.

\subsection{Implications to the coronal heating and solar wind}

Previous observation of the quiet Sun photosphere revealed that the fraction of MBPs involved in merging events, $f_{\rm merge}$, is approximately 0.21 \added{at all times} \citep{Keys_2011_ApJ}.
Furthermore, our analysis has revealed that the mean Poynting flux transported by the magnetic tornado through the transition region, $\mathbf{F_{\rm tornado}}$, is $4.2 \times 10^5 \ \rm erg \ cm^{-2} \ s^{-1}$ \added{(see Figure \ref{fig:sztl})}. This value is approximately four times greater than the \added{mean} Poynting flux generated by granular buffeting, denoted by $\mathbf{F_{\rm buffet}}$, which is $1.0 \times 10^5 \ \rm erg \ cm^{-2} \ s^{-1}$ \added{(see Figure \ref{fig:sztl})}.
Assuming that magnetic tornadoes are produced on every merging MBP, and the amount of the Poynting flux of magnetic tornado and granular buffeting are ubiquitous, we can estimate the contribution of the magnetic tornado to the total Poynting flux to the corona as follows:
\begin{align}
\frac{ f_{\rm merge} F_{\rm tornado} }{ f_{\rm merge} F_{\rm tornado} + (1-f_{\rm merge}) F_{\rm buffet}  } = 0.53,
\end{align}
\noindent i.e., the contribution of magnetic tornadoes is $\approx 50\%$. \added{The contribution may be higher in regions with a stronger coronal magnetic field than that of our simulation ($10$ G), since the filling factor of MCs and thus $f_{\rm merge}$ is expected to be higher in those regions.}
In order to comprehensively investigate the precise contribution of magnetic tornadoes, it is essential to conduct statistical analyses of their occurrence frequency and mean Poynting flux into the corona.

Although our specific target in this study is the coronal loop in the quiet Sun, the similar physical process found in our simulation should occur also in the coronal hole.
The flux imbalance fraction, defined by the net magnetic flux divided by the unsigned magnetic flux, is large ($\ge 0.7$) in coronal holes \citep{Wiegelmann_2004_SolPhys,Zhang_2006_ApJ,Hagenaar_2008_ApJ}, which means that the magnetic patches in a coronal hole tend to exhibit the same polarity.
Since the magnetic tornado is induced by the merger of two MCs with the same polarity, despite small magnetic flux, the coronal hole is also a preferential region for magnetic tornadoes.
The impact of magnetic tornadoes on the solar wind should thus be investigated in future, including its possibility as the origin of magnetic switchbacks \citep{Bale_2019_Nature,Squire_2020_ApJ,Shoda_2021_ApJ}.



\section{Conclusion}
In this work, we investigate the physics of magnetic tornadoes, with a particular focus on energy generation and transfer. To this end, we perform a three-dimensional numerical simulation that self-consistently solves the surface convection and the overlying atmosphere.
A magnetic tornado is triggered by a single merging event of two MCs with the same polarity. \deleted{exhibiting a sign reversal in rotational direction.}
The energy flux carried by the magnetic tornado is four times larger than that caused by granular buffeting, and this is attributed to the reduction of the chromospheric \added{magnetic} energy loss. Our analysis of power spectra suggests that the magnetic energy cascade is weakened in the presence of the magnetic tornado, which may be responsible for the reduction in the energy loss.

\

Numerical computations were carried out on the Cray XC50 at the Center for Computational Astrophysics (CfCA), National Astronomical Observatory of Japan.
M.S. is supported by the Japan Society for the Promotion of Science (JSPS) KAKENHI Grant Number JP22K14077. H.I. is supported by the JSPS KAKENHI Grant Number 19K14756 and the young researcher units for the advancement of new and undeveloped fields of the Institute for Advanced Research (Nagoya University) under the Program for Promoting the Enhancement of Research Universities. T.Y. is supported by the JSPS KAKENHI Grant Number JP21H01124, JP20KK0072, and JP21H04492.

\bibliographystyle{aasjournal}


\listofchanges

\end{document}